\documentclass[journal=jpclcd,manuscript=letter]{achemso}
\usepackage[version=3]{mhchem} % Formula subscripts using \ce{}
\title{A New Class of 3D Topological Insulator in Double Perovskite}

\author{Shu-Ting Pi}
\altaffiliation{Contributed equally to this work}
\author{Hui Wang}
\altaffiliation{Contributed equally to this work}
\author{Jeongwoo Kim}
\author{Ruqian Wu}
\email{wur@uci.edu}
\affiliation{Department of Physics and Astronomy, University of California, Irvine, California 92697-4575, USA }
\author{Yin-Kuo Wang}
\email{kant@ntnu.edu.tw}
\affiliation{Center of General Education, National Taiwan Normal University, Taipei 116, Taiwan}
\author{Chi-Ken Lu}
\email{Lu49@ntnu.edu.tw}
\affiliation{Department of Physics, National Taiwan Normal University, Taipei 11677, Taiwan}

\begin{document}

%\begin{abstract}

\newpage
\begin{abstract}

%\begin{figure}
%\includegraphics[width=0.7\columnwidth]{abstract.pdf}
%\end{figure}

We predict a new class of three-dimensional topological insulators (TIs) in which the spin-orbit coupling (SOC) can more effectively generate band gap. Band gap of conventional TI is mainly limited by two factors, the strength of SOC and, from electronic structure perspective, the band gap when SOC is absent. While the former is an atomic property, the latter can be minimized in a generic rock-salt lattice model in which a stable crossing of bands {\it at} the Fermi level along with band character inversion occurs in the absence of SOC. Thus, large-gap TI's or TI's comprised of lighter elements can be expected. In fact, we find by performing first-principle calculations that the model applies to a class of double perovskites A$_2$BiXO$_6$ (A = Ca, Sr, Ba; X = Br, I) and the band gap is predicted up to 0.55 eV. Besides, surface Dirac cones are robust against the presence of dangling bond at boundary.

%a large
%at $\Gamma$ point
%such as Bi$_2$Se$_3$
% for a range of parameters
% 

%Lastly, in light of the robustness of TI being sensitive to the formation of defect in it and the unintentional doping, the present model could provide a pathway toward large-gap TI with bigger margin of error in material realization. 

%\begin{keyword}
%    Something else
%\end{keyword}
\begin{figure}
\includegraphics[width=\columnwidth]{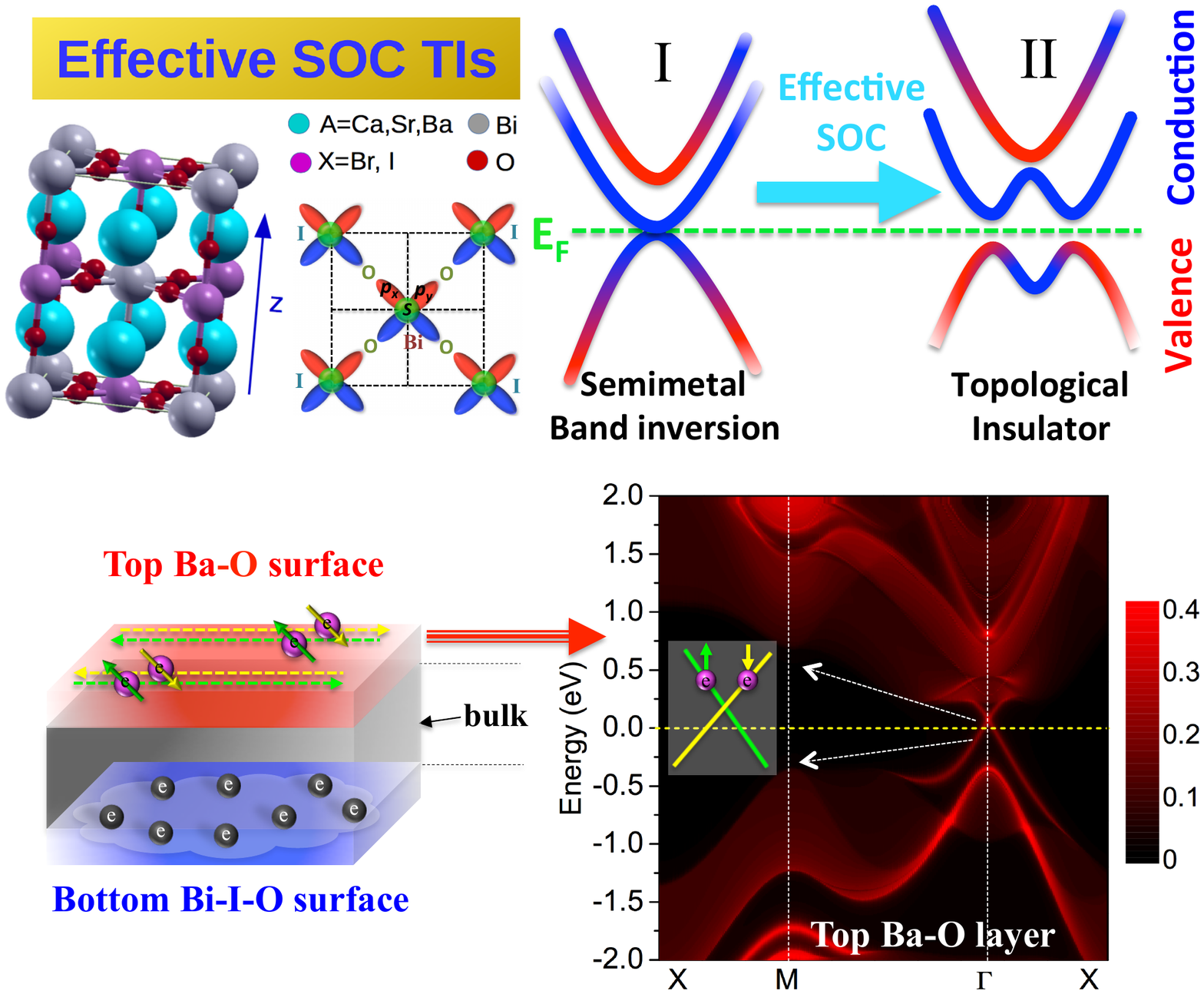}
\end{figure}

\end{abstract}
\maketitle

%\keywords{double perovskite}
\newpage

%{\it Introduction.} 
The discovery of topological phases of matter such as quantum Hall effect~\cite{QHE,Haldane1}, quantum spin Hall effect~\cite{QSH-1,QSH-2,QSH-BHZ,Floquet1,Floquet2,Floquet3}, topolocial crystalline insulators~\cite{TCI}, Dirac semimetals~\cite{DSM-1,DSM-2} and Weyl semimetals~\cite{WSM-1,WSM-2} along with the classifications~\cite{TKNN,class1,classWeyl} is a remarkable success of condensed matter physics and material science. The exotic phenomena rooted in the Dirac electrons on the surface of topological insulators (TIs)~\cite{Majorana1,monopole,FerroDirac,TIexp,TIRev1,YAndoRev} have spurred a bulk of interdisciplinary effort in searching new science and functional materials for practical appliactions~\cite{thermo1,thermo2,thermo3}. 

TIs with large band gaps are particularly desired in achieving room-temperature applications. Tetradymite materials such as Bi$_2$Se$_3$~\cite{Qi_TI} and other combinations~\cite{BiSbTeS} of heavy elements from the V and VI families (see Table I in Ref.~\cite{YAndoRev}) are the prototypical 3D TIs which have demonstrated characteristic behaviors of the surface Dirac fermions~\cite{FerroDirac,TIexp} but at fairly low temperature due to small bulk band gap. Unlike semiconductors, the band gap of TI results from intrinsic spin-orbit coupling (SOC). For 2D materials, the band gap can be increased in a variety of ways, from incorporating heavy adatoms~\cite{adatom}, halogenation~\cite{QSHE1}, fluorination,~\cite{flu} to choosing appropriate substrates~\cite{substrate1,al-substrate} because of the open structure. As for 3D materials, on the other hand, there is little room for tuning the electronic structure and band gap; selecting heavier constituents and/or changing lattice structures seem to be the only feasible route. 

\begin{figure*}[tbp]
\centering
\includegraphics[width=0.9\columnwidth]{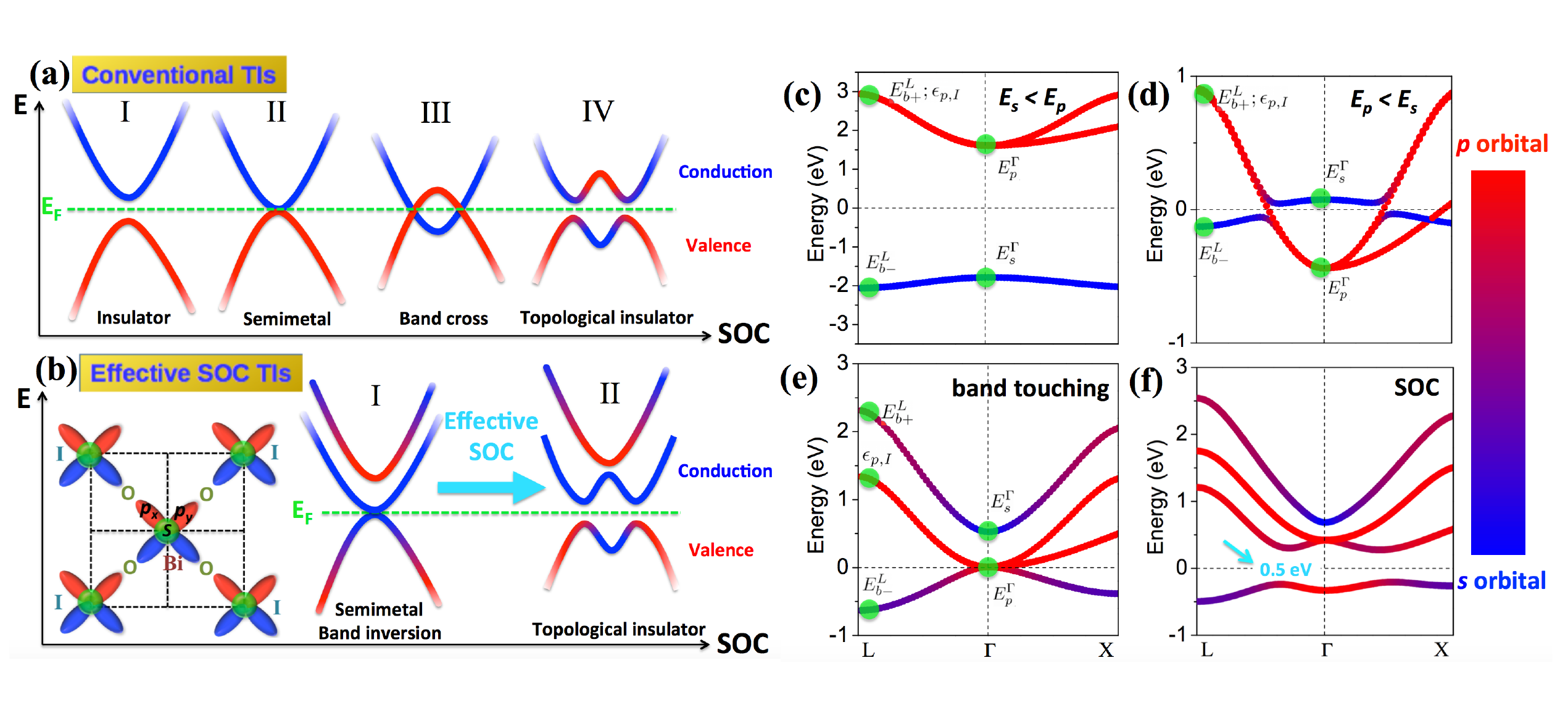}
\caption{(color online) Topological phase transitions that lead to (a) conventional TIs and (b) our proposed TIs due to an {\it effective} SOC. Inset in (b) shows the rock salk structure and its orbitals of our proposed tight-binding model in Eq.~\ref{TB}. (c)-(f) display the bulk spectra by varying the parameters in text. (c) Topological trivial state with the $s$-character Bloch state just below Fermi level at $\Gamma$ point. (d) Band inversion with the $s$-character state above the $p$-character one. (e) Band touching occurs when $E_s^{\Gamma}>E_p^{\Gamma}>E_{b-}^L$ is satisfied. (f) Finally the SOC is turned on and a gap is opened. The Bloch state just below the Fermi level now has the $p$-character and the topologically nontrivial state is arrived.}\label{structure}
\end{figure*}

Figure~\ref{structure}(a) shows the typical evolution of electronic structure during a hypothetical process of turning on the SOC to its full strength. The topological phase transition goes through the three stages: shrinking of band gap (I), gap closing (II), and reemergence of a gap in the topological phase (III and IV)~\cite{QSH-BHZ,TPT}. Therefore, from band structure perspective, the gap in the absence of SOC (stage I) seems to undermine the effectiveness of SOC in generating a large band gap in the topological phase. Here, we propose that the transition in Figure~ \ref{structure}(b) may suggest a different way to make SOC more {\it effective}. Namely, materials possessing band touching at the Fermi level~\cite{YK,QBT,Iridate} in the absence of SOC may become large-gap TIs once the SOC sets in. %Except the prospect of giant-gap TI, the band touching {\it at} the Fermi level are interesting in its own right.

In this work, a tight-binding (TB) model of rock-salt lattice with the unit cell containing a pair of $s$ and $p$ orbitals from two atoms is proposed to implement the novel topological phase transition depicted in Figure~\ref{structure}(b). With such TB model, one may start with a band insulator in which the valence band is of $s$ character and the conduction band is degenerate and of $p$ character. By tuning a set of TB parameters, we are able to achieve the band structure in which the valence and conduction bands, now both having $p$ character, touch {\it at} the Fermi energy, which is protected by the cubic symmetry, and the band of $s$ character lies above them. Consequently, even with a weak SOC, the band touching at the Fermi energy is lifted at once and the highest Bloch state of valence band is of $p$ character. This flip of band character indicates the change of topological nature of the material. 

More importantly, the band touching at the Fermi level and the accompanying band inversion are shown to be universal among the materials reported in this paper. Based on the first-principle calculation and Z$_2$ index evaluation, the double perovskites A$_2$BiXO$_6$ (A = Ca, Sr, Ba; X = I, Br) are TI's and the largest band gap is 0.55 eV, much larger than known {\it pristine} 3D TIs. Further confirmation by employing the Wannier function-based tight-binding calculation allows us to see the Dirac spectrum of the surface states. Since perovskites are stable materials with variable properties by doping, the reported materials could add to the already diverse functionality of perovskites which has been intensively studied.~\cite{DP2,Cao1,Solar2,Solar3,Freeman,Freeman1,Felser,BaTeO}

%{\it Tight-binding model.} 
Band touching in the known materials, graphene for instance, often implies some discrete symmetry from the lattice~\cite{class1}. Here, we introduce a generic tight-binding model for rock-salt lattice from which a band touching at $\Gamma$ point emerges if the band energies at $L$ and $\Gamma$ around the Fermi level are ordered according to Eq.~\ref{inequality}. Here, we consider the Bi and I atoms with each contributing one $s$-orbital and three $p$-orbitals. Dictated by the cubic symmetry, the tight-binding Hamiltonian in momentum space reads 
\begin{align}
H_{TB} = \left(\begin{array}{cccc}
	H_s &  T_x & T_y & T_z\\
	T_x^{\dag} & H_{p_x} & 0 & 0\\
	T_y^{\dag} & 0 & H_{p_y} & 0 \\
	T_z^{\dag} & 0 & 0 & H_{p_z} 
	\end{array}\right) \label{TB}
\end{align} 
 in the representation of the eight atomic orbitals, $[{\rm I}_s, {\rm Bi}_s, {\rm I}_{p_x}, {\rm Bi}_{p_x},{\rm I}_{p_y}, {\rm Bi}_{p_y}, {\rm I}_{p_z}, {\rm Bi}_{p_z}]^T$. The spin is suppressed for the moment, and each element in the matrix is in fact a two-by-two matrix. The diagonal ones read 

\begin{align}
	H_s =  \left(\begin{array}{cccc}
	\epsilon_{s,\rm I} & 2t_{ss}\sum_i\cos\frac{k_ia_i}{2} \\
	2t_{ss}\sum_i\cos\frac{k_ia_i}{2} & \epsilon_{s,\rm Bi}
	\end{array}\right),\label{Hs}
\end{align} and
\begin{align}
	H_{p_x} = \left(\begin{array}{cccc}
	\epsilon_{p,\rm I} &  2\sum_it_{pp}^{\sigma/\pi}\cos\frac{k_ia_i}{2} \\
	 2\sum_it_{pp}^{\sigma/\pi}\cos\frac{k_ia_i}{2} & \epsilon_{p,\rm Bi}
	\end{array}\right),
\end{align}
which are responsible for the coupling between orbitals from different atoms but of the same symmetry. For shorter notation, the sum in the off-diagonal element of $H_{p_x}$ reads $t_{pp}^{\sigma}\cos\frac{k_xa_x}{2}+t_{pp}^{\pi}(\cos\frac{k_ya_y}{2}+\cos\frac{k_za_z}{2})$. Similar expressions in $H_{p_y}$ and $H_{p_z}$ can be inferred from the cubic symmetry. The $s$ and $p$-orbitals are coupled by the off-diagonal matrices, $T_i = (2i)t_{sp}\sin\frac{k_ia_i}{2}  \sigma_x,$ with $i=x,y,z$, and $\sigma_x$ being the first of three Pauli matrices. In the followings, we assume the order of orbital energies: $\epsilon_{s, \rm I}<\epsilon_{s, \rm Bi}<\epsilon_{p, \rm I}<\epsilon_{p, \rm Bi}$. Besides, the Fermi level is above the second lowest band and below the third lowest band of the matrix in Eq.~\ref{TB}.

%\begin{figure}[tbp]
%\centering
%\includegraphics[width=1.0\columnwidth]{TB_band.pdf}
%\caption{ (color online) Bulk spectra of the tight-binding Hamiltonian in Eq.~\ref{TB} by varying the parameters in text. (a) Topological trivial state with the $s$-character Bloch state just below Fermi level at $\Gamma$ point. (b) Band inversion with the $s$-character state above the $p$-character one. (c) Band touching occurs when $E_s^{\Gamma}>E_p^{\Gamma}>E_{b-}^L$ is satisfied. (d) Finally the SOC is turned on and a gap is opened. The Bloch state just below the Fermi level now has the $p$-character and the topologically nontrivial state is arrived.}\label{TB_band}
%\end{figure}
	
At $\Gamma$ point all the matrices $T_{x,y,z}$ vanish identically. Thus, one easily identifies the higher $s$-like state, 
	
\begin{align}
	E_s^{\Gamma} = \frac{\epsilon_{s,\rm Bi}+\epsilon_{s,\rm I}}{2} + \sqrt{
	\left(\frac{\epsilon_{s,\rm Bi}-\epsilon_{s,\rm I}}{2}\right)^2+36t^2_{ss}}  \label{EsG}\:,
\end{align} out of $H_s$ (Eq.~\ref{Hs}), which is close to the Fermi level. Similarly, the lower $p$-like states, 

\begin{align}
	E_p^{\Gamma} =  \frac{\epsilon_{p,\rm Bi}+\epsilon_{p,\rm I}}{2} - \sqrt{
	\left(\frac{\epsilon_{p,\rm Bi}-\epsilon_{p,\rm I}}{2}\right)^2+\left(2t_{pp}^{\sigma}+4t_{pp}^{\pi}\right)^2} \label{EpG}\:,
\end{align} are obtained from $H_{p_{x,y,z}}$ and are triply degenerate due to the cubic symmetry. The states away from these levels are not relevant to the band touching. At the $L$ point, it is less straightforward to obtain the band energies. Nevertheless, we show in the Supplemental Material that the eigenvalues contain two doublets at $\epsilon_{p,\rm I}$ and $\epsilon_{p,\rm Bi}$, respectively, and four nondegenerate states at $E_{a\pm}^L$ and $E_{b\pm}^L$ as a result of $s$-$p$ mixings. It will be clear in a moment that the states, 

\begin{align}
	E_{b\pm}^L = \frac{\epsilon_{s,\rm Bi}+\epsilon_{p,\rm I}}{2} \pm \sqrt{
	\left(\frac{\epsilon_{s,\rm Bi}-\epsilon_{p,\rm I}}{2}\right)^2+12t^2_{sp}
	}\:,\label{EbPM}
\end{align} close to the Fermi level are relevant to the formation of band touching at the $\Gamma$ point. Here, it should be noticed that due to the rock-salt lattice geometry all the hopping parameters $t_{ss}$, $t_{sp}$, and $t_{pp}$ appear in the off-diagonal part of Eq.~\ref{TB} so the band energy is invariant under the mapping $t\longleftrightarrow -t$, the robustness of which is not present in other models.~\cite{Freeman,Felser}

At the outset, the band structure may represent a trivial insulator [Figure~\ref{structure}(c)] in which the higher $s$ band is below the lower $p$ band and the Fermi level lies between them. This situation, $E_s^{\Gamma}<E_p^{\Gamma}$, occurs if the hopping $t_{sp}$ is weak and the on-site energies have $\epsilon_{s, \rm Bi}\ll\epsilon_{p, \rm I}$. As we lift $\epsilon_{s, \rm Bi}$ toward $\epsilon_{p, \rm I}$, the $s$ band at $\Gamma$ can be above the $p$ band [Figure~\ref{structure}(d)] once the condition $E_s^{\Gamma}>E_p^{\Gamma}$ is satisfied. To make a hole band emerge in the presence of the triple degeneracy at $E_p^{\Gamma}$, the only way is to increase $|t_{sp}|$ so that the energy $E_{b-}^L$ is pushed down below $E_p^{\Gamma}$. Finally, the crossing of bands at the Fermi level is established [Figure~\ref{structure}(e)] under the following condition, 

\begin{align}
E_s^{\Gamma}>E_p^{\Gamma}>E_{b-}^L\:,\label{inequality}
\end{align} and, remarkably, the band character is inverted simultaneously. Now the only missing element is a energy gap at the Fermi level so we follow by restoring the spin degree of freedom and turning on the SOC in the form $\lambda\vec L\cdot\vec s$, which leads to the gap opening at $\Gamma$ in Figure~\ref{structure}(f). Aided by the color code in Figure~\ref{structure}(c)-(f) representing the weight of $s$ character of the corresponding Bloch state, one can notice a complete topological phase transition from (c) to (f). Focus on the $\Gamma$ point. Since the Bloch state just below the Fermi level changes from the $s$ character in (c) to the $p$ character in (f), we may conclude that the Z$_2$ invariant~\cite{inversion} associated with Figure~\ref{structure}(c) and (f) must be different and hence the two states must represent distinct phases. In the followings, we shall specifically focus on the double perovskites and show that the above tight-binding model is essential to their band structure and topological property.

Before entering the detailed calculations of real materials, let us pause and compare the TB model in Eq.~\ref{TB} with the well-known BHZ model.~\cite{QSH-BHZ} The latter is a continuum one obtained by projection and has two important ingredients in it, namely the inversion of conduction versus hole bands and the coupling that depends linearly on the crystal momentum. These two crucially depend on the SOC of constituents. In our model, however, it is the tuning of $|t_{sp}|$ which results in the hole band and the band touching at $\Gamma$ point simultaneously. The inversion of band in our model has nothing to do with SOC, either. Moreover, the SOC that couples the electron band with the hole band does not depend on the crystal momentum. Thus, the present model is qualitatively different from the BHZ model.

%Before the end of the section, it is also worthwhile but somewhat elaborate to compare our model with the ABX$_3$ perovskite model proposed in intrinsic~\cite{Freeman} (X=I) and extrinsic~\cite{Felser} (X=O) materials. It may seem that the present double perovskite model with band gap opened at $\Gamma$ is mere a double version of ABX$_3$ model with gap opened at $R$. It The statement is obviously true from the geometry of lattice structure but, as we argue in the following, profoundly invalid in terms of electronic structure. In double perovskite ABB'O$_6$ model, the bands near the Fermi level originate out of the hopping from B to B' or vice versa while those in ABX$_3$ are mainly contributed from B atom. An important consequence of this distinction is that the band energy in the former case is governed by the magnitude of hopping $|t|$ while in the latter case it is controlled by the value of hopping $t$. To be specific, we focus on the valence band of ABX$_3$ in Ref.~\cite{Freeman} and find that the order of energies at $\Gamma$ and $R$ are reversed once the sign of hopping $t_{ss}$ is flipped.~\cite{note2} In our model, on the other hand, the band energy $E_s^{\Gamma}$ in Eq.~\ref{EsG} is not sensitive to the sign of $t_{ss}$. In this sense, our model TI is more stable and has a larger parameter space as well as the possibility to be realized in optical lattice.

\begin{figure}[tbp]
\centering
\includegraphics[width=1.0\columnwidth]{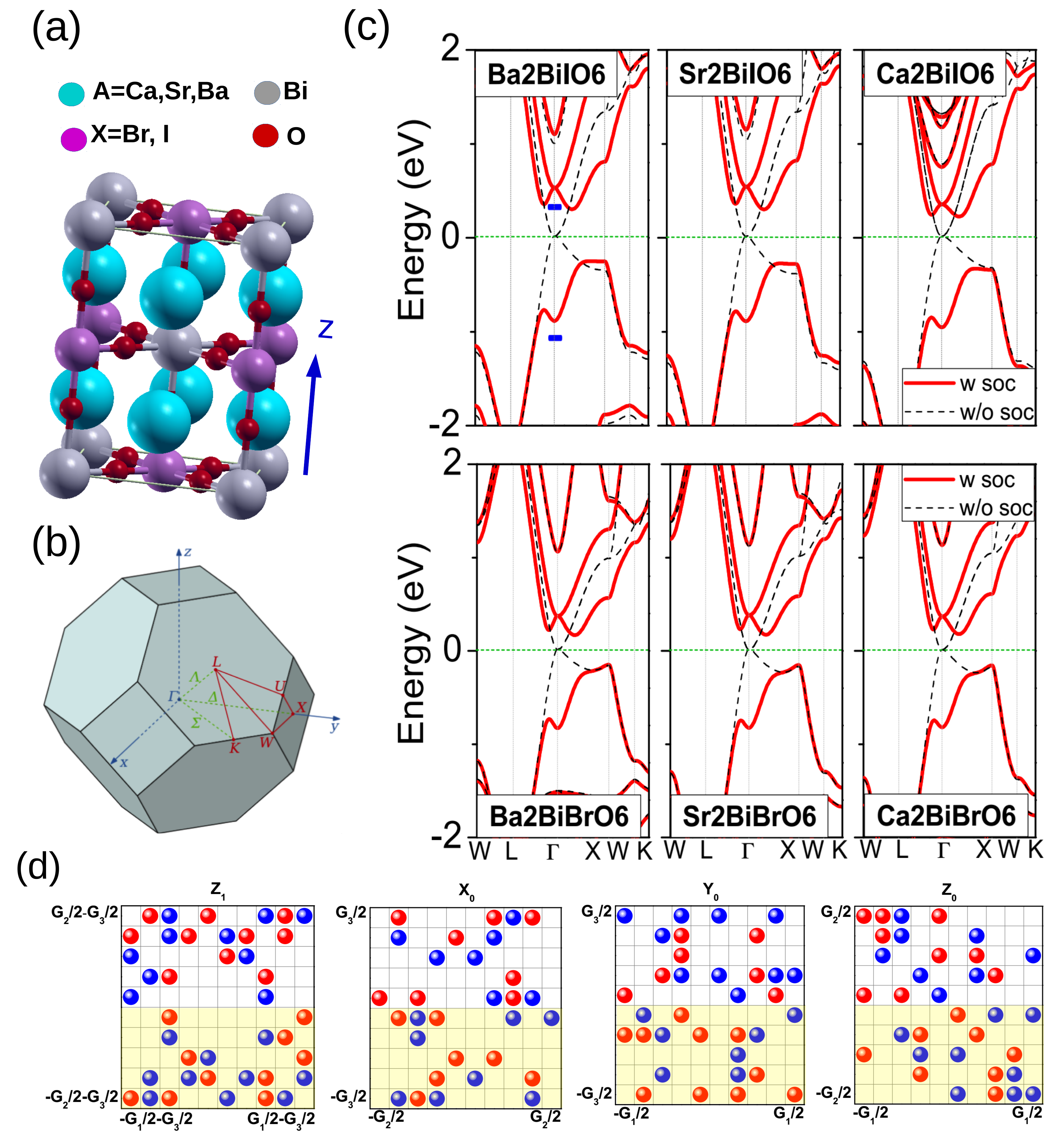}
\caption{ (color online) (a) Crystal structure of double perovskite ABiXO$_6$ (b) Primitive Brillouin zone of double perovskite structure (c) Band structure of various A$_2$BiXO$_6$ materials (d) n-field diagrams of the representative material Ba$_2$BiIO$_6$. The n-field diagrams read $Z_{1}=0$, $Z_{0}=1$, $X_{0}=1$ and $Y_{0}=1$. Using the relation $X_{0}+X_{1}=Y_{0}+Y_{1}=Z_{0}+Z_{1}$ and $(\nu_0;\nu_1 \ \nu_2 \ \nu_3)=(Z_0+Z_1;X_1\ Y_1 \ Z_1 )$ mod 2, we get the Z$_2$ invariant $(\nu_0;\nu_1 \ \nu_2 \ \nu_3)=(1;0\ 0\ 0)$.}\label{band}
\end{figure}

%{\it Band Structure and Z$_2$ Invariant.} 
The class of materials A$_2$BiXO$_6$ in double perovskite structure is shown in the Figure~\ref{band}(a) and its primitive Brillouin zone in Figure~\ref{band}(b). The unit cell contains the octahedra BiO$_6$ and XO$_6$ at the corners of rock-salt lattice and the cation A is surrounded by these octahedra. Determined by the electron negativity of the constituent, the charge state in the ionic limit can be assigned as: A$^{2+}$, Bi$^{3+}$, X$^{5+}$, and O$^{2-}$. Using the first-principles package VASP~\cite{VASP} with generalized gradient approximation (GGA) for the correlation functional~\cite{GGA}, projector augmented wave method (PAW) for core electrons~\cite{PAW}, $10\times 10 \times 10$ for k-sampling integration~\cite{k-sampling} and 500eV for energy cutoff, we find a total of six TIs out of a large number of possible double perovskites (Ca,Sr,Ba)$_2$BB'O$_6$ with B and B' from the III to VII families. The optimized lattice parameters and the band gaps with SOC are listed in Table~\ref{latt_par}. 

\begin{table}[tbp]
\centering
\begin{tabular}{c|| c |c |c } 
 Compound & a=b=c($\AA$) & $\alpha=\gamma=\beta(^\circ)$ & $E_{gap}$ (eV) \\ [0.5ex] 
 \hline\hline
 Ba$_2$BiIO$_6$ & 6.241 & 60 & 0.55  \\
 Sr$_2$BiIO$_6$ & 6.169 & 60 & 0.53  \\
 Ca$_2$BiIO$_6$ & 6.123 & 60 & 0.52  \\
 Ba$_2$BiBrO$_6$ & 6.102 & 60 & 0.33  \\
 Sr$_2$BiBrO$_6$ & 6.009 & 60 & 0.32  \\
 Ca$_2$BiBrO$_6$ & 5.951 & 60 & 0.34  \\ 
\end{tabular}
\caption{The optimized lattice parameters of the primitive cell of A$_2$BiXO$_6$ and their calculated band gaps $E_{gap}$. }\label{latt_par}
\label{table1}
\end{table}

Figure~\ref{band}(c) shows the band structures of the six TIs A$_2$BiXO$_6$. The band structures barely change as the A-site elements vary.  However, the band gap is sensitive to the choice of X. Since the SOC is weaker in Br than in I, the band gap in the materials with Br is only 60$\%$ of those with I. A key feature of these band structures is the band touching at the $\Gamma$-point in the absence of SOC. Clearly, the triple degeneracy is resolved into a doublet and a singlet in the presence of SOC, which is consistent with the previous TB model.

% Figure 3
\begin{figure}[tbp]
\centering
\includegraphics[width=1.0\columnwidth]{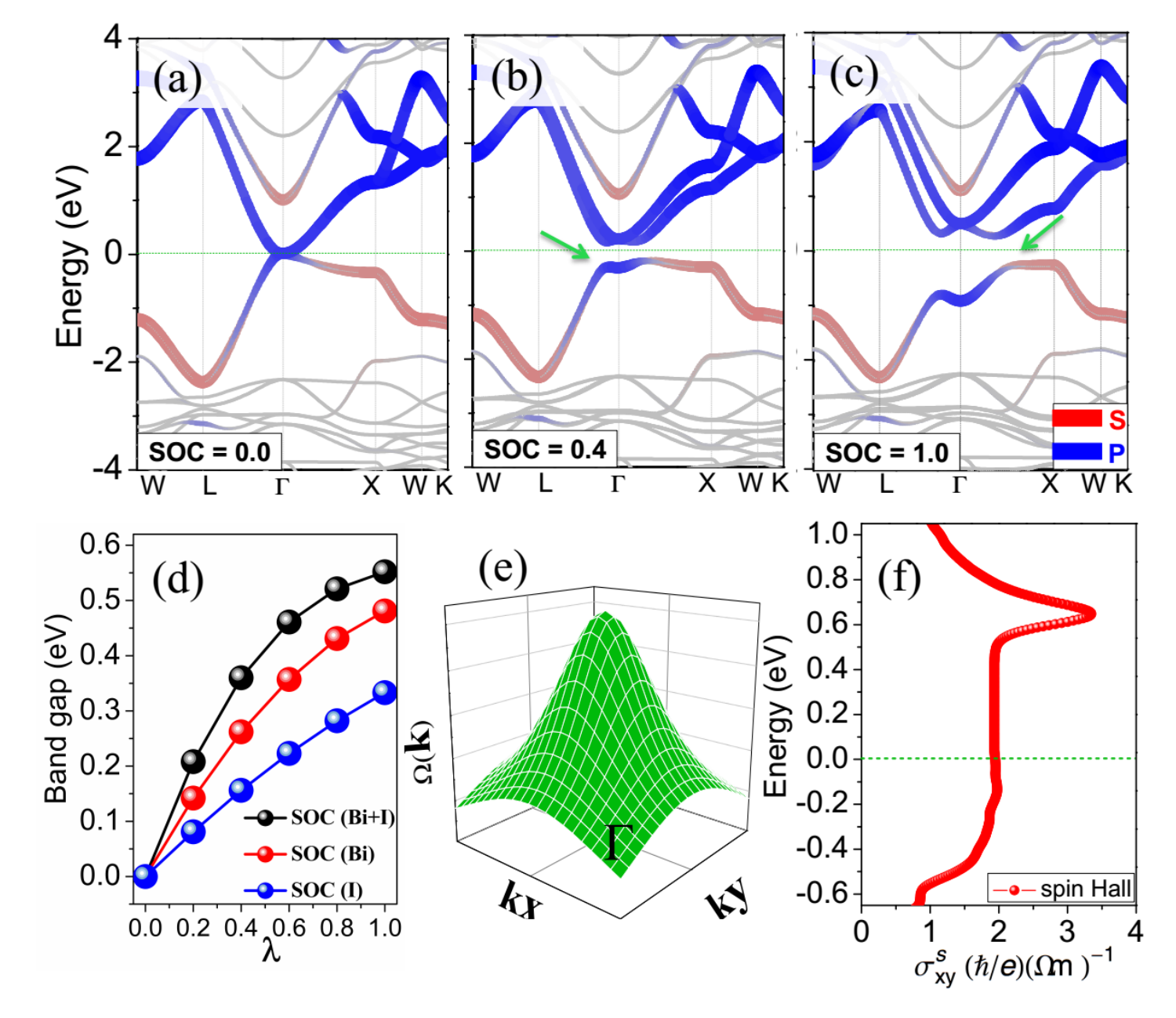}
\caption{ (color online) Evolution of band structure of Ba$_2$BiIO$_6$ as SOC is gradually turned on [(a) to (c)]. Colored intensity shows the component of s-orbital and p-orbital on Bi and I. In (a), the band inversion at $\Gamma$ clearly appears even if SOC is turned off. Arrows indicate the k-point where narrowest gaps appear. Inset shows the band gap while tuning the SOC. (d) shows the band gaps when tuning the strength of SOC of Bi and I respectively. (e) displays the calculated spin Berry curvature around $\Gamma$-point and (f) the spin Hall conductance as a function of Fermi energy.}\label{fatband}
\end{figure}

To examine the topological feature of these materials, we implement the n-field method of evaluating the Z$_2$ invariant proposed by Fukui et. al.~\cite{n-field1,n-field2}. The resultant n-field diagrams of a representative material Ba$_2$BiIO$_6$ are shown in Figure~\ref{band}(d). We found that the six materials share the same Z$_2$ invariants ($\nu_0$; $\nu_1$ $\nu_2$ $\nu_3$) = (1; 0 0 0), indicating that all the six materials are strong TIs. On the other hand, Fu and Kane proposed another method to obtain the Z$_2$ invariant for crystals with inversion symmetry by evaluating the parity eigenvalues of the occupied bands~\cite{inversion}. Table~\ref{table1} displays the invariant $\delta(\Lambda_i)$ associated with the full set of occupied bands of Ba$_2$BiIO$_6$ at the eight time-reversal invariant momenta (TRIM) $\Lambda_i$ as well as the parity eigenvalues $\xi_v$ ($\xi_c$) of the band just below (above) the Fermi level. We thus obtain the same topological invariant ($\nu_0$; $\nu_1$ $\nu_2$ $\nu_3$) = (1; 0 0 0), consistent with the n-field method. 
  
% Table I
\begin{table}[tbp]
\centering
\begin{tabular}{c|| c c c c c c c c} 
 $\Lambda_i$ & $\Lambda_{000}$ & $\Lambda_{100}$ & $\Lambda_{010}$ & $\Lambda_{110}$ & $\Lambda_{001}$ & $\Lambda_{101}$ & $\Lambda_{011}$ & $\Lambda_{111}$ \\ [0.5ex] 
 \hline\hline
 $\xi_v$ & -- & -- & -- & + & -- & + & + & -- \\
 $\xi_c$ & -- & + & + & -- & + & -- & -- & + \\
 $\delta(\Lambda_i)$ & -- & -- & -- & + & -- & + & + & -- \\ 
\end{tabular}
\caption{The topological invariant $\delta$ associated with the occupied bands at eight TRIM's $\Lambda_i$ and the parity eigenvalues of the Bloch states just below ($\xi_v$) and above ($\xi_c$) the Fermi level.}
\label{table1}
\end{table}

Next, we continue with studying the band characters near the Fermi level. The respective weight of $s$- and $p$-components of the Bloch states are calculated by DFT and are shown by the color code in Figure~\ref{fatband} in which the SOC is gradually turned on to its full strength. In the case of vanishing SOC [Figure~\ref{fatband}(a)], the lowest conduction band is composed of $p$-orbitals only while the highest valence band has the main component from $s$-orbitals except near the $\Gamma$ point. From the aspect of Z$_2$ invariant, one may regard this state as a TI with zero band gap. According to Figure~\ref{fatband}(b) and (c), as the SOC is gradually turned on, the size of band gap increases accordingly while the band character remains, eventually leading to a large band gap [Figure~\ref{fatband}(d)]. It is important to note that the SOC in this class of materials {\it does not} serve to exchange the band character, or the parity eigenvalue, unlike what it does in conventional TI Bi$_2$Se$_3$. To further clearly identify the non-trivial topology, we also calculate the spin Berry curvature [Figure~\ref{fatband}(e)] associated with the occupied band and the resultant spin Hall conductance [Figure~\ref{fatband}(f)] as a function of Fermi energy. The quantized values of spin Hall conductance within the SOC gap demonstrates the critical features of the topological phase. 

%The spin Berry curvature significantly different at gamma points ascribed to the discontinuities of $s-$ and $p-$orbitals.

% Figure 4
\begin{figure}[tbp]
\centering
\includegraphics[width=1.0\columnwidth]{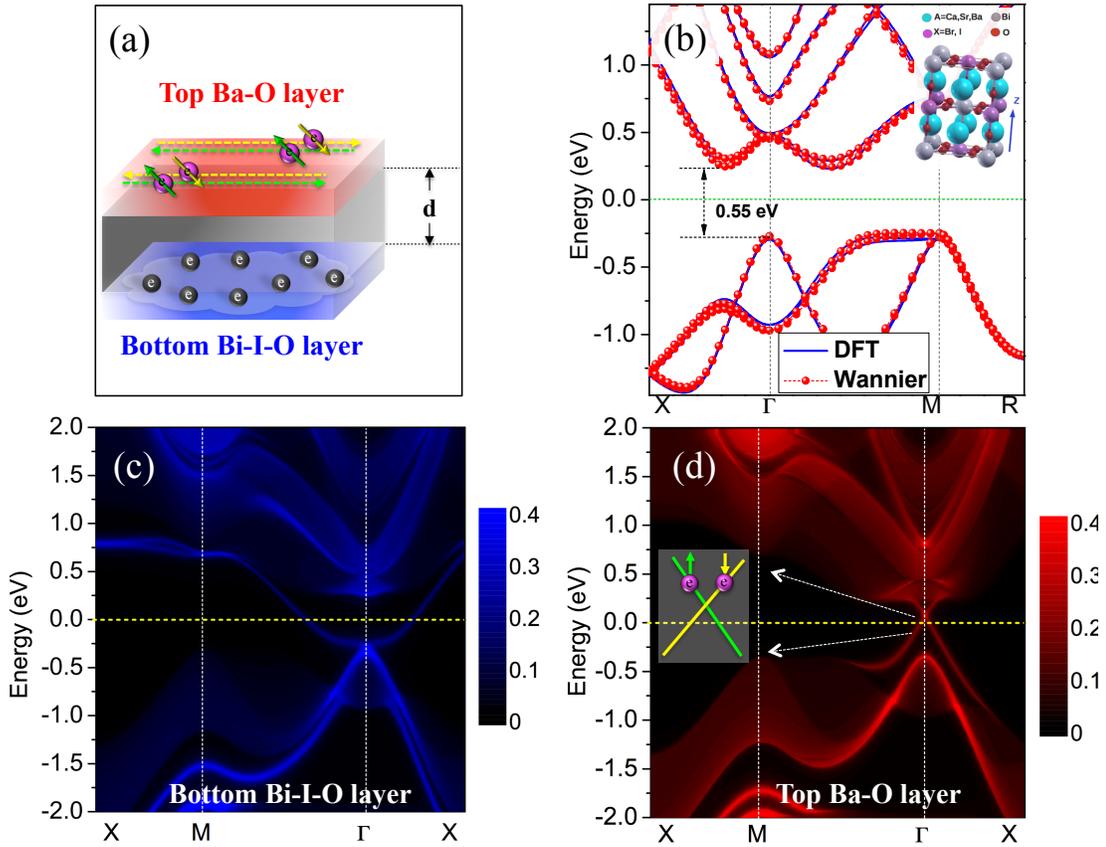}
\caption{ (color online) Surface band structure of Ba$_2$BiIO$_6$ using Wannier functions.
(a) slab geometry in our calculation. For Wannier calculation, thickness d=$\infty$; for first-principles calculation, d=4nm. (b) displays the bulk band structure from Wannier functions. (c) spectrum of the bottom surface (Bi-O-I layer termination). (d)spectrum of the top surface (Ba-O layer termination.}\label{sband}
\end{figure}

%Figure 5
\begin{figure}[tbp]
\centering
\includegraphics[width=1.0\columnwidth]{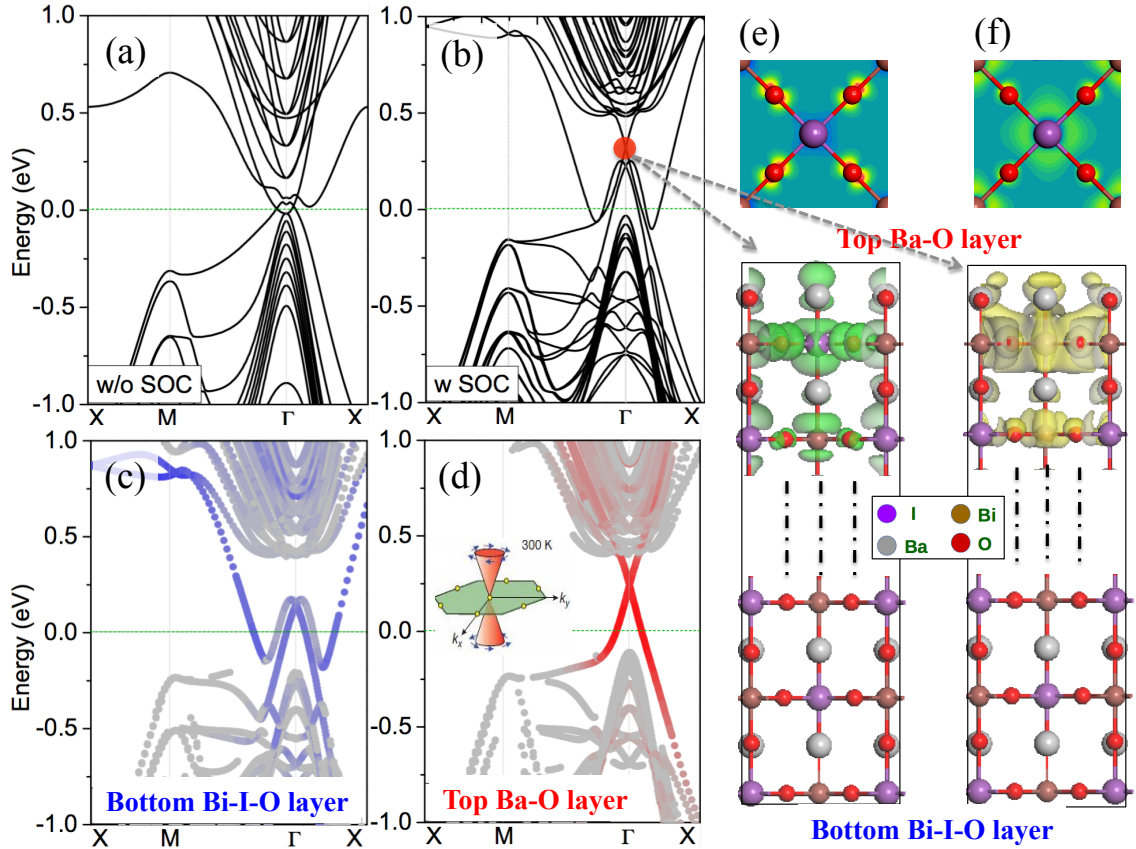}
\caption{ (color online) Surface band structure of Ba$_2$BiIO$_6$ using DFT calculation.
(a) and (b) the full band structure of the slab geometry without and with SOC, respectively. (c) projected spectrum on the bottom surface (Bi-O-I layer). (d) projected spectrum on the top surface (top second Bi-O-I layer). Inset shows the helical state feature. (e) and (f) the spatial wave functions of the two states at the Dirac points circled in (b).}\label{swave}
\end{figure}

%{\it Surface states.} 
Another benchmark of a TI is the presence of linear dispersion within the bulk gap representing the Dirac electron on the surface. To investigate the surface states, we consider the semi-infinite slab structure with two types of surface. The one at top is terminated at the Ba-O layer while the one at bottom is at Bi-I-O layer, which is shown in Figure~\ref{sband}(a). A tight-binding Hamiltonian is constructed with the maximally localized Wannier functions (MLWFs) using WANNIER90 package \cite{W90}. The bulk band structures from MLWF are shown in Figure~\ref{sband}(b), showing excellent agreement comparing to DFT bands. The surface spectral weight of the semi-infinite slab is computed by the surface Green'€™s function method \cite{SGF-1,SGF-2,SGF-3}. Surprisingly, as shown in Figure~\ref{sband}(c), a non-Dirac-like dispersion appears at the bottom Bi-I-O layer at which dangling bonds present. Since the Bloch states around the Fermi level have significant components falling on Bi and I atoms, the presence of dangling bond may greatly alter the wavefunctions and the surface states. On the other hand, as shown in Figure~\ref{sband}(d), a Dirac-like spectrum is clearly presented on the top surface terminated by Ba-O layer since the Bloch states associated with Ba are away from the Fermi energy. It is worthy mentioning that the Ba-O layer may be regarded as a cap layer that protects the Dirac state from the severe perturbation due to the dangling bond. 

Although Wannier functions method is commonly used to explore the surface states,~\cite{Qi_TI} it is usually considered as an artificial calculation because it uses the tight-binding parameters obtained from the bulk band structure without taking the interface effects such as surface energy renormalization and dangling bonds into account. To address this issue, we also perform the DFT calculation of a slab geometry in which 5 unit cells of Figure~\ref{band}(a) are repeated along the z-axis. It corresponds to the thickness about 4 nm, which has been proven large enough against the interactions between top and bottom surfaces based on our Wannier function projected bands of analysis. The band structures of the slab geometry without and with SOC are shown in Figure~\ref{swave}(a) and (b), respectively. In Figure~\ref{swave}(c) for the bottom surface, a distorted in-gap Dirac spectrum state emerges around the $\Gamma$-point. This distorted Dirac surface states actually resemble those appear on the surface of Bi$_2$Te$_3$ [see Figure~8(b) in Ref.~\cite{YAndoRev}]. Instead, the Dirac state on the top surface [Figure~\ref{swave}(d)] is more transparent and consistent with the Wannier-function analysis in previous paragraph. According to our TB model and DFT calculation, the bulk states around the Fermi level are mainly contributed by the $s$- and $p$-orbitals of Bi and I atoms. When the bottom Bi-O-I layer is exposed, the dangling bonds certainly alter the states around the Fermi level, which might result in the distortion of the Dirac state. As for the case with Ba-O layer at the termination, the associated dangling bonds can only perturb the states away from the Fermi level, leaving the surface Dirac states and the Kramers doublets intact. Hence we expect the class of materials with surfaces terminated at A-O layers to possess the topological features and superior surface transport property. In fact, the Ba-O layer termination is also more structurally stable because the breaking of BiO$_6$ or IO$_6$ octahedra requires more energy, making the top Ba-O surface more feasible during the synthesis process. We have also plotted the 3D spatial wave functions of the pair of Dirac states [circled in Figure~\ref{swave}(b)] in Figure~\ref{swave}(e) and (f) which shows a highly localized behavior on the top second Bi-O-I layer. The 2D projections of surface wave functions clearly exhibit the $s$- and $p$-characters also in agreement with our previous discussion. 

%Because the Ba-O layers do not contribute to the states around the Fermi level, we select the Bi-O-I layers at the top and at the bottom for the projection of surface spectrum.
%{\it Discussion.} 
While the presence of robust surface Dirac states crucially distinguishes TI from the ordinary semiconductor, there could be another aspect that might differentiate the two classes from computational perspective. Namely, the GGA-PBE functionals often underestimates the band gap of semiconductors.~\cite{HSE1} An improvement is to adopt hybrid functional such as HSE,~\cite{HSE2} which employs a screened exchange interaction while keeping the long-ranged Coulomb interaction. It is worthwhile to stress that SOC, a relativistic effect, is the main factor in determining the band gap of the predicted TI's. Since SOC has little to do with Hatree-Fock interaction, we believe that the adoption of HSE is irrelevant to the band gap of TI.    

%it is worthwhile to 

One distinct feature of our proposed 3D TIs is that it is intrinsic TIs with the Fermi level lying in the large non-trivial SOC gap. Synthesis of the predicted material, Ba$_2$BiIO$_6$ for instance, is an interesting yet challenging direction for chemists and material scientists. Currently, the compound BaBiO$_3$ is a known superconducting material~\cite{BaBiO3} while the study of solid phase of iodine oxide~\cite{Iodine} is rare although the iodine chemistry is of importance to the research of earth ozone.~\cite{Iodine1} Besides, the predicted TI shall show robustness against disorder. For double perovskite A$_2$BB'O$_6$, the well-known disorder includes the self-doping, the intermixing between B and B', and the oxygen deficiency. From our model, the intermixing may be more substantially disadvantageous since the Bloch states near the Fermi level are contributed by the B and B' orbitals, but it can be overcome experimentally in a more controlable synthesis process. To consider another realistic situation, the robustness of TI states against surface reconstruction is investigated. As shown in Figure S1, one can see that surface relaxation mainly induces charge redistribution between the bottom and top surface from the shifting of Fermi level, leaving the main features of Dirac states unchanged. Experimentally, one can detect the topological surface states by measuring the quantized conductance protected by the topological nature, or by mapping out the surface state of solids in the reciprocal space by the state-of-art Angle-resolved photoemission spectroscopy (ARPES). Since the energy position of these surface states can be easily tuned by applying electric field, one may switch between electron and hole transport on the surfaces as well. Moreover, doping to double perovskite structure can also be implemented by replacing the A$^{2+}$ cations with A$^{1+}$ or A$^{3+}$ ones of similar size~\cite{DP2}, which makes the Bloch states near the Fermi level intact while little disorder is introduced to the system. Thus, the doped compounds can be converted into topological superconductor~\cite{Ong,Fu1} at sufficiently low temperature, which has immense impact on the fundamental physics of non-Abelian particles and fault-tolerant quantum computations~\cite{Ivanov}. We believe that our work may provide an interesting platform and stimulate new experimental investigations of double-pervoskite structures in many other new directions.

%{\it Summary.} 
By proposing a generic tight-binding model for rock-salt lattice, we have shown a new notion of topological phase transition from a trivial state in Figure~\ref{structure}(c) to the topological state in (f) of 3D materials with strong SOC and cubic symmetry. With the band touching and band inversion in the band structure without SOC, it turns out that the SOC becomes more effective in generating a large band gap of a TI. We further suggest possible realization of our model in well-known double perovskites A$_2$BiXO$_6$ (A=Ca, Sr, Ba; X=I, Br) by doping V and VII elements into B and B$^{'}$ site, with non-trivial band gap as large as 0.55eV. Further analysis on the surface states shows that the surface Dirac cone is stable on surface terminated by A and oxygen atoms and robust against surface reconstruction, providing stable TIs states for experimental realization. Our findings not only enrich the physics of TIs realm but also open a new vista of searching for new TIs in new dimensions. Our proposed materials hold great potential to realize a new paragon of 3D TIs for practical applications.  

%\section{acknowledgement}
Useful conversation with M. Kennett is acknowledged. Work at National Taiwan Normal University was supported by Taiwan Ministry of Science and Technology through Grant No. 103-2112-M-003-012-MY3 and No. 103-2112-M-003-005. Work at University of California, Irvine was supported by DOE-BES (Grants No. DE-FG02-05ER46237 for H.W. and No. SC0012670 for S.P.). Computer simulations were partially supported by NERSC and National Center for High-Performance Computing of Taiwan.

Supporting Information. Derivation of the band energy in Eq.~\ref{EbPM} and surface band structure considering the presence of dangling bond included. 

%\section{Author contributions}
%YKW performed preliminary search of the materials. CKL developed the tight-binding model. HW, STP and JWK performed the analysis of the electronic structutre and topological properties. CKL, STP and HW discussed the results and wrote the maniscript. RQW, CKL and YKW supervised the project.    


\begin{thebibliography}{plain}

% QHE
\bibitem{QHE} von Klitzing, K.; Dorda, G; Pepper, M. New Method for High-Accuracy Determination of the Fine-structure Constant Based on Quantized Hall Resistance. \emph{Phys. Rev. Lett.} {\bf 1980}, {\it 45}, 494.  

\bibitem{Haldane1} Haldane, F.~D.~M. Model for a Quantum Hall Effect without Landau Levels: Condensed-matter Realization of the ``Parity Anomaly". \emph{Phys. Rev. Lett.} {\bf 1988}, {\it 61}, 2015.
% QSHE

\bibitem{QSH-1} Kane, C.~L.; Mele, E. J. Quantum Spin Hall Effect in Graphene. \emph{Phys. Rev. Lett.} {\bf 2005}, {\it 95}, 226801.
\bibitem{QSH-2} Bernevig, B. A.; Zhang, S.-C. Quantum Spin Hall Effect. \emph{Phys. Rev. Lett.} {\bf 2006}, {\it 96}, 106802. 
\bibitem{QSH-BHZ} Bernevig, B. A.; Hughes T.~L.; Zhang, S.-C. Quantum Spin Hall Effect and Topological Phase Transition in HgTe Quantum Wells. \emph{Science} {\bf 2006}, {\it 314}, 1757-1761. 

\bibitem{Floquet1} Oka, T.; Aoki, H. Photovoltaic Hall Effect in Graphene, \emph{Phys. Rev. B} {\bf 2009}, {\it 79}, 106802. 

\bibitem{Floquet2} Linder, N.~H.; Rafel, G.; Galitski, V. Floquet Topological Insulator in Semiconductor Quantum Wells, \emph{Nat. Phys.} {\bf 2011}, {\it 7}, 490-495.

\bibitem{Floquet3} Pi, S.-T.; Savrasov, S.~Y. Polarization Induced Z$_2$ and Chern Topological Phases in a Periodically Driving Field. \emph{Sci. Rep.} {\bf 2016}, {\it 6}, 22993. 

% TCI
\bibitem{TCI} Fu, L. Topological Crystalline Insulators. \emph{Phys. Rev. Lett.} {\bf 2011}, {\it 106}, 106802. 

% DSM
\bibitem{DSM-1} Wang, Z.; Sun, Y.; Chen, X.-Q.; Franchini, C.; Xu, G.; Weng, H.; Dai, X.; Fang, Z. Dirac Semimetal and Topological Phase Transitions in A$_3$Bi (A=Na, K, Rb). \emph{Phys. Rev. B} {\bf 2012}, {\it 85}, 195320.

\bibitem{DSM-2} Wang, Z.; Weng, H.; Wu, Q.; Dai, X.; Fang, Z. Three-dimensional Dirac Semimetal and Quantum Transport in Cd$_3$As$_2$. \emph{Phys. Rev. B} {\bf 2013}, {\it 88}, 125427.
 
% WSM
\bibitem{WSM-1} Wan, X.; Turner, A.~M.; Vishwanath, A.; Savrasov, S.~Y. Topological Semimetal and Fermi-arc Surface States in the Electronic Structure of Pyrochlore Iridates. \emph{Phys. Rev. B} {\bf 2011}, {\it 83}, 205101.

\bibitem{WSM-2} Xu, S.-Y.; Belopolski, I.; Alidoust, N.; Neupane, M.; Bian, G.; Zhang, C.; Sankar, R.; Chang, G.; Yuan, Z.; Lee, C.-C.; {\it et al.} Discovery of a Weyl Fermion Semimetal and Topological Fermi Arcs. \emph{Science} {\bf 2015}, {\it 349}, 613-617.

%Huang, S.-M.; Zheng, H.; Ma. J.; Sanchez, D.~S.; Wang, B.; Bansil, A.; Chou, F.; Shibayev, P.~P.; Lin, H.; Jia, S.; Hasan, M.~Z.

% classification
\bibitem{TKNN} Thouless, D.~J.; Kohmoto, M.; Nightingale, M.~P.; den Nijs, M. Quantized Hall Conductance in a Two-dimensional Periodic Potential. \emph{Phys. Rev. Lett.} {\bf 1982}, {\it 49}, 405.

\bibitem{class1} Schnyder, A.~P.; Ryu, S.; Furusaki, A; Ludwig, A.~W.~W. Classification of Topological Insulators and Superconductors in Three Spatial Dimensions. \emph{Phys. Rev. B} {\bf 2008}, {\it 78}, 195125. 

\bibitem{classWeyl} Yang B.-J.; Nagaosa, N. Classification of Stable Three-dimensional Dirac Semimetals with Nontrivial Topology. \emph{Nat. Commun.} {\bf 2014}, {\it 5}, 4898.

% superconductivity
\bibitem{Majorana1} Fu L.; Kane, C.~L. Superconducting Proximity Effect and Majorana Fermions at the Surface of a Topological Insulator. \emph{Phys. Rev. Lett.} {\bf 2008}, {\it 100}, 096407. 

% monopole
\bibitem{monopole} Qi, X.-L.; Li, R.; Zang, J.; Zhang, S.-C. Inducing a Magnetic Monopole with Topological Surface States. \emph{Science} {\bf 2009}, {\it 323}, 1184-1187. 

% exp
\bibitem{FerroDirac} Checkelsky, J.~G.; Ye, J.; Onose, Y.; Iwasa, Y.; Tokura, Y. Dirac-fermion-mediated Ferromagnetism in a Topological Insulator. \emph{Nat. Phys.} {\bf 2012}, {\it 8}, 729-733.

\bibitem{TIexp} Zhang, J.; Chang, C.-Z.; Tang, P.; Zhang, Z.; Feng, X.; Li, K.; Wang, L.-L.; Chen, X.; Liu, C.; Duan, W.; {\it et al.} Topology-driven Magnetic Quantum Phase Transition in Topological Insulators. \emph{Science} {\bf 2013}, {\it 339}, 1582-1586.

%; He, K.; Xue, Q.-K.; Ma, X.; Wang, Y.

% TI Review
\bibitem{TIRev1} Hasan, M.~Z.; Kane, C.~L. {\it Colloquium}: Topological Insulators. \emph{Rev. Mod. Phys.} {\bf 2010}, {\it 82}, 3045. 

\bibitem{YAndoRev} Ando, Y. Topological Insulator Materials. \emph{J. Phys. Soc. Jpn.} {\bf 2013}, {\it 82}, 102001.


% thermo
\bibitem{thermo1} Hor, Y. S.; Richardella, A.; Roushan, P.; Xia, Y.; Checkelsky, J.~G.; Yazdani, A.; Hasan, M.~Z.; Ong, N.~P.; Cava, R.~J. $p$-type Bi$_2$Se$_3$ for Topological Insulator and Low-temperature Thermoelectric Applications. \emph{Phys. Rev. B} {\bf 2009}, {\it 79}, 195208.

\bibitem{thermo2} Xu, Y.; Gan, Z.; Zhang, S.-C. Enhanced Thermoelectric Performance and Anomalous Seebeck Effects in Topological Insulators. \emph{Phys. Rev. Lett.} {\bf 2014}, {\it 112}, 226801.

\bibitem{thermo3} Pal, K.; Anand, S.; Waghmare, U.~V. Thermoelectric Properties of Materials with Nontrivial Electronic Topology. \emph{J. Mater. Chem. C} {\bf 2015}, {\it 3}, 12130-12139.

% Bi2Se3

\bibitem{Qi_TI} Zhang, H.; Liu, C.-X.; Qi, X.-L.; Dai, X.; Fang, Z.; Zhang, S.-C. Topological Insulator in Bi$_2$Se$_3$, Bi$_2$Te$_3$, and Sb$_2$Te$_3$ with a Single Dirac Cone on the Surface. \emph{Nat. Phys.} {\bf 2009}, {\it 5}, 438-442. 

% BiSbTeS
\bibitem{BiSbTeS} Kushwaha, S.~K.; Pletikosic, I.; Liang, T.; Gyenis, A.; Lapidus, S.~H.; Tian, Y.; Zhao, H.; Burch, K.~S.; Lin, J.; Wang, W.; {\it et al.} Sn-doped Bi$_{1.1}$Sb$_{0.9}$Te$_2$S Bulk Crystal Topological Insulator with Excellent Properties. \emph{ Nat. Commun.} {\bf 2016}, {\it 7}, 11456.

%; Ji, H.; Fedorov, A.~V.; Yazdani, A.; Ong, N.~P.; Valla, T.; Cava, R.~J.

% adatom & substrate

\bibitem{adatom} Hu, J.; Alicea, J.; Wu, R.~Q.; Franz, M. Giant Topological Insulator Gap in Graphene with 5$d$ Adatoms. \emph{Phys. Rev. Lett.} {\bf 2012}, {\it 109}, 266801. 

% halogenation
\bibitem{QSHE1} Ma, Y.; Dai, Y.; Niu, C.; Huang, B. Halogenated Two-dimensional Germanium: Candidate Materials for Being of Quantum Spin Hall State. \emph{J. Mater. Chem.} {\bf 2012}, {\it 22}, 12587-12591.

% fluorination

\bibitem{flu} Xu, Y.; Yan, B.; Zhang, H.-J.; Wang, J.; Xu, G.; Tang, P.; Duan, W.; Zhang, S.-C. Large-gap Quantum Spin Hall Insulators in Tin Films. \emph{ Phys. Rev. Lett.} {\bf 2013}, {\it 111}, 136804.

%substrate
\bibitem{substrate1} Zhou, M.; Ming, W.; Liu, Z.; Li, P.; Liu, F. Epitaxial Growth of Large-gap Quantum Spin Hall Insulator on Semiconductor Surface. \emph{PNAS} {\bf 2014}, {\it 111}, 14378-14381.

\bibitem{al-substrate} Wang, H.; Pi, S.-T.; Kim, J.; Wang, Z.; Fu, H.~H.; Wu, R.~Q. Possibility of Realizing Quantum Spin Hall Effect at Room Temperature in Stanene/Al$_2$O$_3$(0001). \emph{Phys. Rev. B} {\bf 2016}, {\it 94}, 035112.

% phase transition

\bibitem{TPT} Murakami, S. Phase Transition Between the Quantum Spin Hall and Insulator Phases in 3D: Emergence of a Topological Gapless Phase. \emph{New J. Phys.} {\bf 2007}, {\it 9}, 356.

% band touching

\bibitem{YK} Vafek O.; Yang, K. Many-body Instability of Coulomb Interacting Bilayer Graphene: Renormalization Group Approach. \emph{Phys. Rev. B} {\bf 2009}, {\it 81}, 041401(R). 

\bibitem{QBT} Sun, S.; Yao, H.; Fradkin, E.; Kivelson, S.~A. Topological Insulators and Nematic Phases from Spontaneous Symmetry Breaking in 2D Fermi Systems with a Quadratic Band Crossing. \emph{Phys. Rev. Lett.} {\bf 2009}, {\it 103}, 046811.

\bibitem{Iridate} Yamaji Y.; Imada, M. Metallic Interface Emerging at Magnetic Domain Wall of Antiferromagnetic Insulator: Fate of Extinct Weyl Electrons. \emph{Phys. Rev. X} {\bf 2014}, {\it 4}, 021035.

% Perovskite

\bibitem{DP2} Vasala S.; Karppinen, M. A$_2$B'B"O$_6$ Perovskites: A Review. \emph{Prog. Solid State Chem.} {\bf 2015}, {\it 43}, 1-36.

\bibitem{Cao1} Kim, B.~J.; Jin, H.; Moon, S.~J.; Kim, J.-Y.; Park, B.-G.; Leem, C.~S.; Yu, J.; Noh, T.~W.; Kim, C.; Oh, S.-J.; {\it et al.} Novel $J_{{\rm eff}}$ = 1/2 Mott State Induced by Relativistic Spin-orbit Coupling in Sr$_2$IrO$_4$. \emph{Phys. Rev. Lett.} {\bf 2008}, {\it 101}, 076402. 

%; Park, J.-H.; Durairaj, V.; Cao, G.; Rotenberg, E.

\bibitem{Solar2} Kojima, A.; Teshima, K.; Shirai, Y.; Miyasaka, T. Organometal Halide Perovskites as Visible-light Sensitizers for Photovoltaic Cells. \emph{J. Am. Chem. Soc.} {\bf 2009}, {\it 131}, 6050-6051. 

\bibitem{Solar3} Burschka, J.; Pellet, N.; Moon, S.-J.; Humphry-Baker, R.; Gao, P.; Nazeeruddin, M.~K.; Gratzel, M. Sequential Deposition as a Route to High-performance Perovskite-sensitized Solar Cells. \emph{Nature} {\bf 2013}, {\it 499}, 316-319.

\bibitem{Freeman} Jin, H.; Im, J.; Freeman, A.~J. Topological Insulator Phase in Halide Perovskite Structures. \emph{Phys. Rev. B} {\bf 2012}, {\it 86}, 121102(R).

\bibitem{Freeman1} Jin, H.; Rhim, S.~H.; Im, J.; Freeman, A.~J. Topological Oxide Insulator in Cubic Perovskite Structures. \emph{ Sci. Rep.} {\bf 2013}, {\it 3}, 1651.

\bibitem{Felser} Yan, B.; Jasen, M.; Felser, C. A Large-energy-gap Oxide Topological Insulator Based on the Superconductor BaBiO$_3$. \emph{Nat. Phys.} {\bf 2103}, {\it 9}, 709-711.

\bibitem{BaTeO} Yang, M.; Wang, R.~N. Topological Insulator in Tellurium-based Perovskites, \emph{ Int. J. Mod. Phys. B} {\bf 2015}, {\it 29}, 1550073.

%\bibitem{note2} Please also see Figure S1 in Supplemental Information for clarification.

%methods 
\bibitem{VASP} Kresse, G.; Hafner, J. {\it Ab initio} Molecular-dynamics Simulation of the Liquid-metal-amorphous-semiconductor Transition in Germanium. \emph{Phys. Rev. B} {\bf 1994}, {\it 49}, 14251.
\bibitem{GGA} Perdew, J. P.; Burke, K.;  Ernzerhof, M. Generalized Gradient Approximation Made Simple. \emph{Phys. Rev. Lett.} {\bf 1996}, {\it 77}, 3865.
\bibitem{PAW} Kresse, G.; Joubert, D. From Ultrasoft Pseudopotentials to the Projector Augmented-wave Method. \emph{Phys. Rev. B} {\bf 1999}, 59, 1758.; Blochl, P.~E. Projector augmented-wave method. \emph{Phys. Rev. B} {\bf 1994}, {\it 50}, 17953.
\bibitem{k-sampling} Monkhorst, H.~J.; Pack, J.~D. Special Points for Brillouin-zone Integrations. \emph{Phys. Rev. B} {\bf 1976}, {\it 13}, 5188. 

% method
\bibitem{n-field1} Fukui, T.; Hatsugai, Y.;  Suzuki, H. Chern Numbers in Discretized Brillouin Zone: Efficient Method of Computing (Spin) Hall Conductances. \emph{J. Phys. Soc. Jap.} {\bf 2005}, {\it 74}, 1674-1677. 
\bibitem{n-field2} Fukui T.; Hatsugai, Y. Quantum Spin Hall Effect in Three Dimensional Materials: Lattice Computation of Z$_2$ Topological Invariants and Its Application to Bi and Sb. \emph{J. Phys. Soc. Jap.} {\bf 2007}, {\it 76}, 053702.
\bibitem{inversion} Fu, L.; Kane, C.~L. Topological Insulators with Inversion Symmetry. \emph{Phys. Rev. B} {\bf 2007}, {\it 76}, 045302.

% Wannier

\bibitem{W90} Mostofi, A. A.; Yates, J. R.; Pizzi, G.; Lee, Y.~S.; Souza, I.; Vanderbilt, D.;  Marzari, N. An Updated Version of Wannier90: A Tool for Obtaining Maximally-Localised Wannier Functions. \emph{Comput. Phys. Commun.} {\bf 2014}, {\it 185}, 2309-2310.
\bibitem{SGF-1} Sancho, M. P. L.; Sancho, J. M. L.; Rubio, J. Quick Iterative Scheme for the Calculation of Transfer Matrices: Application to Mo (100). \emph{J. Phys. F: Met. Phys.} {\bf 1984}, {\it 14}, 1205-1215.  
\bibitem{SGF-2} Sancho, M. P. L.; Sancho, J. M. L.; Sancho, J. M. L.; Rubio, J. Highly Convergent Schemes for the Calculation of Bulk and Surface Green Functions. \emph{J. Phys. F: Met. Phys.} {\bf 1985}, {\it 15}, 851.
\bibitem{SGF-3} Lee P.; Kim, J.; Kim, J.; Ryu, M.; Park, H.; Kim, N.; Kim, Y.; Lee, N.; Kioussis, N.; Jhi, S.; {\it et al.} Topological Modification of the Electronic Structure by Bi-bilayers Lying Deep Inside Bulk Bi$_2$Se$_3$. \emph{J. Phys.: Condens. Matter} {\bf 2016}, {\it 28}, 085002.

%\bibitem{footnote} 
% HSE

\bibitem{HSE1} Zhao, Y.; Truhlar, D.~G. Calculation of Semiconductor Band Gaps with the M06-L Density Functional, \emph{J. Chem. Phys.} {\bf 2009}, {\it 130}, 074103.

\bibitem{HSE2} Heyd, J.; Scuseria, G.~E.; Ernzerhof, M. Hybrid Functional Based on a Screened Coulomb Potential, \emph{J. Chem. Phys.} {\bf 2004}, {\it 118}, 8207-8215.  

\bibitem{BaBiO3} Sleight, A. W.; Gillson, J. L.; Bierstedt, P. E. High Temperature Superconductivity in the BaPb$_{1-x}$Bi$_x$O$_3$ Systems. \emph{ Solid State Commun. } {\bf 1975}, {\it 17}, 27-28.

\bibitem{Iodine} Nikitin, I.~V. Halogen Monoxides, \emph{Russ. Chem. Rev.} {\bf 2008}, {\it 77}, 739-749.

\bibitem{Iodine1} Saiz-Lopez, A.; Fernandez, R.~P.; Ordonez, C.; Kinnisons, D.~E.; Gomez Martin, J.~C.; Lamarque, J.-F.; Tilmes, S. Iodine Chemistry in the Troposphere and its Effect on Ozone, \emph{ Atoms. Chem. Phys. } {\bf 2014}, {\it 14}, 13119-13143. 
% topological SC

\bibitem{Ong} Hor, Y.~S.; Williams, A~.J.; Checkelsky, J.~G.; Roushan, P.; Seo, J.; Xu, Q.; Zandbergen, H.~W.; Yazdani, A.; Ong, N.~P.; Cava, R.~J. Superconductivity in Cu$_x$Bi$_2$Se$_3$ and its Implications for Pairing in the Undoped Topological Insulator. \emph{Phys. Rev. Lett.} {\bf 2010}, {\it 104}, 057001.

\bibitem{Fu1} Fu, L.; Berg, E. Odd-Parity Topological Superconductors: Theory and Application to Cu$_x$Bi$_2$Se$_3$. \emph{Phys. Rev. Lett.} {\bf 2010}, {\it 105}, 0097001.

\bibitem{Ivanov} Ivanov, D.~A. Non-Abelian Statistics of Half-Quantum Vortices in $p$-Wave Superconductors. \emph{Phys. Rev. Lett.} {\bf 2001}, {\it 86}, 268.

\end{thebibliography}
\end{document}